\documentclass[twocolumn,floatfix,showpacs,preprintnumbers,amsmath,amssymb,aps,prl]{revtex4}
\usepackage{amssymb,amsmath,graphics,epsfig,color}
\begin{document}

\title{A Bright, Slow Cryogenic Molecular Beam Source for Free Radicals}
\author{J.F. Barry, E.S. Shuman, and D. DeMille}
\affiliation{Department of Physics, Yale University, PO Box 208120, New Haven, CT 06520, USA}
\date{\today}

\begin{abstract}
We demonstrate and characterize a cryogenic buffer gas-cooled molecular beam source capable of producing bright beams of free radicals and refractory species. Details of the beam properties (brightness, forward velocity distribution, transverse velocity spread, rotational and vibrational temperatures) are measured under varying conditions for the molecular species SrF. Under typical conditions we produce a beam of brightness $1.2 \times 10^{11}$ molecules/sr/pulse in the $X^2\Sigma^+ (v=0,N_{rot}=0)$ state, with $140 \frac{m}{s}$ forward velocity and a rotational temperature of $\approx \!$ 1 K. This source compares favorably to other methods for producing beams of free radicals and refractory species for many types of experiments. We provide details of construction that may be helpful for others attempting to use this method.

\end{abstract}

\maketitle

There has recently been a renewed interest in diatomic molecules because of their potential applications to a variety of areas.  In particular, cold polar molecules are expected to be useful in fields including quantum computation \cite{DeMille2002,Maxwell2006}, quantum simulation of condensed matter systems \cite{Goral2002,Micheli2006,Barnett2006}, precision measurements \cite{Vutha2010,Leanhardt2010,Hudson2002,Cahn2008}, and controlled chemistry \cite{Balakrishnan2001,Krems2008,Hudson2006a}. Although many methods for cooling and trapping molecules have been proposed \cite{Carr2009}, only a few have been demonstrated. One successful technique for producing cold molecules is by binding laser-cooled atoms into molecules via photoassociation \cite{Sage2005} or Feshbach resonance \cite{Ni2008}; however, both these methods have so far been limited to bialkali molecules.

A more general technique for trapping molecules is to directly load them into a trap from a molecular beam \cite{DeMille2004}.  In this case the number of molecules trapped is directly tied to the flux, translational and internal (rotational and vibrational) temperatures, and forward velocity of the beam.  Specifically, slow, cold molecular beams with high flux are ideal for loading traps.  Such beams of molecules are also useful for high precision spectroscopy \cite{Flambaum2007}.

Unfortunately, traditional supersonic and effusive molecular beams are not ideally suited to these requirements.  Thermal effusive sources can produce translationally slow molecules, but the finite rovibrational temperatures of the molecules vastly reduce the effective flux of these beams.  On the other hand, supersonic beam sources are quite efficient at cooling both internal and external molecular degrees of freedom, but the resultant molecular beam velocity is quite high, which prohibits direct molecular trapping.

The development of general methods for directly producing slow, cold molecules applicable to a variety of molecular species remains an active area of research.  Tremendous advances have been made in the deceleration of supersonic beams in the past decade.  Stark deceleration \cite{Meerakker2005,Hudson2006}, Zeeman deceleration \cite{Vanhaecke2007,Narevicius2008}, optical deceleration \cite{Fulton2004}, collisional deceleration \cite{Chandler2003}, and rotating nozzles \cite{Gupta1999,Strebel2010} have all been developed to slow supersonic beams to velocities of a few m/s, where molecules can be loaded into a trap.  Although this is a highly general technique for producing beams of slow, cold molecules, it becomes significantly more challenging for unstable molecules (e.g. free radicals) or molecules that do not have substantial vapor pressure at room temperature.  The high forward velocity in supersonic beams also makes slowing heavy molecules much more challenging, and thus far the slowing techniques have primarily been limited to light molecules \cite{Tarbutt2004}.  Furthermore, these slowing techniques have so far resulted in molecular beams with relatively low flux.

Another promising technique is cryogenic buffer gas gas cooling.  Typically in these experiments, the target molecules and the buffer gas are enclosed in a cell. Collisions with the buffer gas cool both the internal and external degrees of freedom to the temperature of the cell.  This technique has been demonstrated for a wide variety of molecular species \cite{Krems2009}, with resulting molecular temperatures as low as 100 mK and the ability to trap molecules demonstrated in situ \cite{Weinstein98}. The primary disadvantage of buffer gas cooling is that the cold molecules remain in the cell where collisions prohibit most further measurements and manipulation of the molecules.  It remains a significant challenge to separate the buffer gas from the molecules of interest.

Cryogenic buffer gas cooling has also served as the basis for molecular beams \cite{Maxwell2005,Buuren2009,Patterson2007,Patterson2009}.  In these experiments, the cell has an exit aperture through which the molecules can escape and form a beam.  These sources can be operated both in an effusive regime and a ``hydrodynamically enhanced'' regime, depending on the buffer gas density.  For low buffer gas densities, there are relatively few collisions near the exit aperture, and the beam exits effusively from the cell.  Effusive beams are characterized by thermal molecular beam velocities and low flux due to poor extraction of molecules from the cell.  For higher buffer gas densities, hydrodynamic effects inside the cell lead to nearly complete extraction of molecules from the cell.  However, in the hydrodynamic regime, collisions between molecules and buffer gas near the exit aperture can result in a boosted forward velocity.

Here we report on the development and characterization of an ablation-loaded, cryogenic buffer gas beam source of a diatomic free radical. The source is designed to produce a beam of molecules which will subsequently be laser cooled \cite{Shuman2009,Shuman2010}, and ultimately loaded into a trap.  We operate the source at intermediate buffer gas densities, where the extraction efficiency of molecules into the beam is high, while the forward velocity of the beam is significantly lower than the full supersonic speed of the buffer gas.  Previous studies on such beams have focused on only a few of the properties under a limited range of conditions \cite{Maxwell2005,Buuren2009,Patterson2007,Patterson2009}.  The characterization given here provides a more complete description of a cryogenic buffer gas beam source, including measurement of the beam brightness, forward velocity distribution, transverse velocity spread, and rotational and vibrational temperatures over a wide range of buffer gas densities.  We also characterize the molecular beam flux into a room-temperature apparatus for the first time.  We find that the brightness of the cryogenic beam source compares very favorably with other sources capable of producing beams of similar refractory species.  Although the measurements presented here were all conducted using strontium monofluoride (SrF), we expect similar performance from other species which can be vaporized by ablation of solid precursors.

The remainder of the paper is structured as follows: Part II details basic properties of buffer gas beams. Part III describes the experimental construction and design principles.  Part IV presents measurements of the beam properties, additional observations regarding source operation and a detailed comparison to existing sources based on other technologies. Part V consists of a short conclusion.

\section{General Source Properties}
The basic principles of a buffer gas beam source are simple. A cold cell is held at temperature $T_0$ while gaseous buffer gas atoms $b$, also at $T_0$, are continuously flowed through the cell at a rate $\mathcal{F}$. The value of $\mathcal{F}$ can typically be continuously varied over a wide range of values, which allows control over the density of $b$ inside the cell. The target molecules $a$ are injected into the cell at an initial temperature $T_a(t=0)\gg T_0$, in our case by laser ablation of a solid precursor inside the cell. These initially hot molecules undergo many collisions with the buffer gas, which cool both the translational and rotational degrees of freedom to near $T_0$. During and after this thermalization, the target molecules diffuse through the buffer gas to the cell walls, where they stick and are lost. Simultaneously, both the target molecules and the buffer gas are extracted into a beam through a hole in the cell.  The ratio of the time scales for these competing processes determines the efficiency of molecule extraction from the cell.  Meanwhile the number of collisions between the buffer gas atoms and the molecules around the exit aperture determines the divergence, forward velocity distribution, and internal temperatures of the molecules in the beam.  Although these basic principles of buffer gas beam sources are simple, the actual dynamics in and around the cell can be quite complex for ablation-loaded sources such as ours.  In the remainder of this section we describe relevant characteristics of the cell conditions and the resulting molecular beam.  The description given here is meant to provide a qualitative description of a few of the relevant parameters of these sources.  More sophisticated models would be necessary to give a complete quantitative description.

\subsection{Mean Free Path}
The basic properties of the buffer gas beam are primarily determined by the mean free paths $\lambda_a$ and $\lambda_b$ of particles $a$ and $b$ inside the cell. Under typical operating conditions, the density $n_b$ of the buffer gas far exceeds that of the target molecules $n_a$.  This allows two simplifying assumptions to be made. First, collisions involving two $a$ particles are rare and therefore may be ignored; thus $\lambda_a$ then depends only on $n_b$. Similarly, collisions between two $b$ particles are much more likely than between $a$ and $b$ particles, so $\lambda_b$ also depends only on $n_b$.  Under these conditions the mean free paths are
\begin{equation}\label{eq:meanfreepath}
\lambda_a = \frac{1}{n_b\sigma_{ab}\sqrt{m_a/m_b+1}} \hspace{5 mm} \text{and} \hspace{5 mm}
\lambda_b = \frac{1}{{\sqrt{2}n_b\sigma_{bb}}},
\end{equation} where $\sigma_{ab}$ is the elastic collision cross section between species $a$ and $b$, $\sigma_{bb}$ is the elastic collision cross section between $b$ particles, and $m_a$ and $m_b$ are the masses of $a$ and $b$, respectively.

In practice $b$ is usually a noble gas, so $n_b$ is not easy to measure using laser absorption or fluorescence techniques.  Instead the value of $n_b$ in our experiments is inferred from simple cell dynamics described here.  Under steady state conditions, the flow rate of $b$ into the cell, $\mathcal{F}$, will equal the rate of $b$ out.  The rate at which particles of $a(b)$ are emitted out of the source exit aperture (with area $A$) into solid angle $d\omega$ at angle $\theta$ is
\begin{equation}\label{eq:ramsey}
dQ_{a(b)} = \frac{d\omega}{4\pi} n_{a(b)} v^{exit}_{a(b)} A \cos \theta,
\end{equation} where $v^{exit}_{a(b)}$ is the mean velocity at the exit aperture of species $a(b)$ \cite{Ramsey1956}.

The values of $v^{exit}_a$ and $v^{exit}_b$ can vary significantly depending on the number of collisions particles of the given species experience near the exit aperture.  The Reynolds number for $a$ and $b$, defined as $Re_a = \frac{d}{\lambda_a}$ and $Re_b = \frac{d}{\lambda_b}$ where $d$ is the cell aperture diameter, characterizes the number of collisions each species experiences while exiting the cell aperture. Typically $m_a > m_b$ and $\sigma_{ab} \approx \sigma_{bb}$, so we assume hereon that $Re_a > Re_b$. For $1 \gg Re_a > Re_b$, the molecules exit the hole effusively.  For $Re_a > Re_b \gg 1$, the particles undergo many collisions around the exit aperture, resulting in supersonic velocities as the beam escapes the source.

We relate $n_b$ to the flow rate $\mathcal{F}$ via the following reasoning. Under effusive conditions, $v^{exit}_b$ is the same as the mean velocity of $b$ inside the cell: $v^{exit}_b=\bar{v}_b=(2/\sqrt{\pi})\beta$.  Here $\beta \equiv \sqrt{2 k_B T_b / m_b}$, where $T_b$ is the translational temperature of $b$.  Integration of Eq. (\ref{eq:ramsey}) over all angles leads to a total rate of $Q_b^e = \frac{n_b A \beta}{2\sqrt{\pi}}$, in the effusive regime.  In the fully supersonic regime, the value of $v^{exit}_b$ is less clear.  Collisions in and around the aperture boost the forward velocity of the buffer gas atoms from $\bar{v}_b$ up to a maximum value given by the fully supersonic velocity \cite{Scoles88,Pauly2000}
\begin{equation}\label{eq:supersonicvelocity}
 v_{b\parallel}^{s}=\sqrt{\gamma/(\gamma-1)}\beta,
\end{equation}
where for a noble gas $\gamma = 5/3$. We can obtain an upper limit on the total rate in the supersonic regime by assuming that all particles of $b$ exit the cell along the beam line at $v_{b\parallel}^{s}$.  Under these assumptions Eq. (\ref{eq:ramsey}) yields an upper limit on the total rate $Q^s_b = \sqrt{5/2} n_b A \beta$. By equating $\mathcal{F}$ to $Q_b$ we arrive at the density of $n_b$ for the two cases given by
\begin{equation}\label{eq:n}
n_b = \frac{\kappa \mathcal{F}}{A \beta},
\end{equation}
where $\kappa=\kappa^{e}=2\sqrt{\pi}$ for fully effusive and $\kappa=\kappa^{s} = 1/\sqrt{\gamma/(\gamma-1)}$ for fully supersonic; hence $\frac{\kappa^{e}\mathcal{F}}{A \beta}>n_b>\frac{\kappa^{s} \mathcal{F}}{A \beta}$.

\subsection{Thermalization}

The extraction of target molecules into the beam depends critically on their thermalization with the buffer gas.  If the molecules do not thermalize before they make contact with the cell walls, they will stick and be lost.  Thermalization of initially hot molecules inside a cold buffer gas cell can be described by a simple kinematic model \cite{Decarvalho1999}. At time $t=0$, ablation creates $N_a^{cell}$ particles of species $a$ at initial high translational temperature $T_a(0)$. The particles cool via collisions with the buffer gas; the translational temperature $T_a(\mathcal{N})$ of species $a$ after $\mathcal{N}$ collisions can be written as a differential equation:
\begin{equation}\label{eq:cool}
\frac{dT_a(\mathcal{N})}{d\mathcal{N}} = -\frac{T_a(\mathcal{N})-T_b}{C},
\end{equation}
where $C \equiv (m_a+m_b)^2/(2m_am_b)$. This simple model assumes $T_b=T_0$ at all times. Integration of Eq. (\ref{eq:cool}) yields
\begin{equation}\label{eq:temp}
T_a(\mathcal{N})=T_b+(T_a(0)-T_b) e^{-\mathcal{N}/C}.
\end{equation}
The temperature of species $a$ then asymptotically approaches $T_b$.  We define a nominal number of collisions necessary to thermalize species $a$, $\mathcal{N}_{Th}$, as the number of collisions such that $T_a(\mathcal{N}_{Th})$ is of the same order of magnitude as $T_b$, i.e. by setting the second term in Eq. (\ref{eq:temp}) equal to $T_b$.  This yields $\mathcal{N}_{Th}\equiv C~\textrm{log}\left(\frac{T_a(0)-T_b}{T_b}\right)\approx C~\textrm{log}\left( \frac{T_a(0)}{T_b} \right)$.  The resulting thermalization time $\tau_{Th}$ is given by
\begin{equation}\label{eq:thermalizationtime}
\tau_{Th}= \frac{\mathcal{N}_{Th}}{R},
\end{equation}
where $R \approx n_b \sigma_{ab} \bar{v}_b \sqrt{1+m_b/m_a} $ is the approximate collision rate for species $a$.

The value of $\mathcal{N}_{Th}$ allows us to estimate the minimum density $n_b$ necessary to achieve thermalization.  After $\mathcal{N}_{Th}$ collisions, each particle $a$ travels a characteristic distance $X_{Th}$, which must range between the distance traveled by a diffusive random walk and purely ballistic flow, or $\sqrt{\mathcal{N}_{Th}}\lambda_a\leq X_{Th}\leq\mathcal{N}_{Th}\lambda_a$.  The heavy and initially hot molecules have much more momentum than the light and cold buffer gas, so we take the ballistic limit $X_{Th} = \mathcal{N}_{Th}\lambda_a$. For ablation in the middle of a cubic cell of side length $L_c$, the particles of $a$ exiting through the aperture will be efficiently thermalized only if $X_{Th}\lesssim L_c/2$.  Using Eq. (\ref{eq:meanfreepath}) leads to a density requirement for thermalization given by
\begin{equation}\label{eq:thermalizationdensity}
n_{Th} \gtrsim \frac{2 C}{\sigma_{ab} L_c \sqrt{m_a/m_b + 1}}\textrm{log}\left(\frac{T_a(0)-T_b}{T_b}\right).
\end{equation}
For most species $\sigma_{ab}$ is unknown; however, $\sigma_{ab}$ typically shows little variation among target species $a$ for a given buffer gas $b$ \cite{Lu2009,Nguyen2005,Tsikata2010,Jessiephd2007,Skoff2010}, and for noble gases $\sigma_{ab} \approx \sigma_{bb}$~\cite{Hamel1986,Hogervorst1971}.  Furthermore, the exact value of $T_a(0)$ is unknown, and must be estimated  (typical estimates for ablation temperatures are $T_a(0) \sim 10^4$ K \cite{Davis1985}).  Nonetheless, we expect that Eq. (\ref{eq:thermalizationdensity}) provides the correct order of magnitude for the density required for thermalization.

Similar arguments are applicable to the internal (vibrational and rotational) temperatures of $a$, provided that the appropriate collisional cross sections are used.  In general the collision cross sections for vibrational, rotational and translational relaxation obey \cite{Scoles88,Pauly2000}
\begin{equation}\label{eq:cross sections}
\sigma^{vib}_{ab} \ll \sigma^{rot}_{ab} < \sigma_{ab},
\end{equation}
where $\sigma^{vib}_{ab}$, $\sigma^{rot}_{ab}$, and $\sigma_{ab}$ are the respective vibrational, rotational and translational collisional cross sections between species $a$ and buffer gas $b$.  In this case we expect similar thermalization behavior for rotational and translational degrees of freedom, while vibrational thermalization may occur over much longer time scales.

\subsection{Diffusion and Extraction}
After thermalization, particles of both species $a$ and $b$ are extracted from the cell through the cell aperture and into the beam.  The efficiency of extraction of $a$ through the hole is primarily limited by the diffusion of these particles to the cell walls.  The diffusion of species $a$ into species $b$ at temperature $T$ is governed by the diffusion equation, $\frac{dn_a}{dt} = \nabla^2(Dn_a)$~\cite{Hasted1972}.  Here $D$ is the diffusion constant, given to good approximation by $D = 3/(16\sigma_{ab} n_b)\times\sqrt{2 \pi k_B T/\mu}$~\cite{Hasted1972}, where $\mu = (m_a m_b)/(m_a+m_b)$ is the reduced mass.  We can therefore approximate the time for species $a$ to be lost to the cell walls via diffusion, $\tau_{diff}$, by the time constant of the lowest-order diffusion mode \cite{Hasted1972}, giving
\begin{equation}\label{eq:diffusion}
\tau_{diff} \approx \frac{L_c^2}{4\pi^2 D}.
\end{equation}
We approximate $\tau^b_{pump}$, the time constant governing the extraction of $b$ through the cell aperture, by the typical time for the cell volume to be emptied by flow out of the exit aperture:
\begin{equation}\label{eq:pumpout}
\tau^b_{pump} \approx \frac{L_c^3 n_b}{\mathcal{F}}=\frac{\kappa L_c^3}{A \beta}.
\end{equation}
In the remainder of the paper we assume that species $a$ is fully entrained in the flow of species $b$ inside the cell.  With this assumption, species $a$ also will exit the cell with the same time constant. The cell extraction efficiency $\epsilon$ is the fraction of $a$, which, once produced and thermalized inside the cell, is extracted into a beam. The quantity
\begin{equation}\label{eq:efficiency}
\xi \equiv \frac{\tau_{diff}}{\tau^b_{pump}}\propto\frac{\mathcal{F}}{L_c}
\end{equation}
has been found to be strongly correlated with $\epsilon$~\cite{Patterson2007}.  Eq. (\ref{eq:efficiency}) suggests that small cells operated at high flow rates are ideal for maximal extraction efficiencies. For $\xi \gg 1$ we expect particles of $a$ to exit the cell before they diffuse to the cell walls; therefore in this ``hydrodynamic'' regime we expect to observe $\epsilon_{hyd} \sim 1$. Values of $\epsilon > 0.4$ have been reported \cite{Patterson2007} for $\xi \gtrsim 1$. For $\xi \ll 1$ we expect purely diffusive in-cell behavior, with $\epsilon$ determined by the geometric extraction efficiency $\sim A/(\pi L_c^2)$ for molecules produced in the center of the cell. In this regime values of $\epsilon \sim 0.001$ have been reported \cite{Maxwell2005}.

\subsection{Beam Formation}
As the molecules pass through the exit aperture, the number of collisions that particles of $a$ and $b$ experience determines to a large extent the properties of the beam.  For $1 \gg Re_a > Re_b$ (effusive regime), there are no collisions for either species in the vicinity of the aperture and the extracted beam is purely effusive. In this regime the mean forward velocities of $a$ and $b$ in the beam, denoted by $v_{a\parallel}$ and $v_{b\parallel}$ respectively, are given by \cite{Ramsey1956}
\begin{equation}\label{eq:effusivevelocity}
v_{a\parallel} = v_{a\parallel}^e=\frac{3}{4}\sqrt{\pi}\alpha ~~~\text{and}~~~
v_{b\parallel} = v_{b\parallel}^e=\frac{3}{4}\sqrt{\pi}\beta,
\end{equation} where $\alpha \equiv \sqrt{2 k_B T_a/m_a}$.
We also expect the translational temperatures of $a$ and $b$ in the beam obey $T_{a}^{beam} = T_b^{beam} = T_0$.

For $Re_a > Re_b \gg 1$ (supersonic regime), all particles experience many collisions as they exit the aperture and expand into vacuum. Because $n_b\gg n_a$ the buffer gas species $b$ drives the expansion, and the properties of $b$ in the beam determine to a large extent the beam properties of $a$. During the isentropic expansion, $v_{b\parallel}$ increases while $T_b^{beam}$ cools, resulting in a boosted but narrow velocity distribution. Using a simple hard-sphere scattering model to describe the cooling during the expansion \cite{Pauly2000}, we can estimate $T_b^{beam}$ to be
\begin{equation}\label{eq:supersonicfinaltemp}
T_b^{beam} \lesssim 3.12(\sigma_{bb} n_b d)^{-\frac{4}{5}}T_0.
\end{equation}
This value of $T_b^{beam}$ represents an upper bound on the beam temperature because it neglects quantum mechanical effects which become more important for low values of $T_b^{beam}$ \cite{Toennies1977,Wang1988}, particularly for $b=$ He. In the supersonic regime we expect $T_a^{beam}$ to approach $T_b^{beam}$. We also expect $v_{a\parallel}$ to approach $v_{b\parallel}$ and $v_{b\parallel}$ to approach $v_{b\parallel}^s$.

\section{Experimental Apparatus}

In this experiment, SrF is the molecule of interest $a$, and the buffer gas species $b$ is He.  The apparatus is built around a 2-stage closed cycle pulse tube refrigerator (Cryomech PT415). A vacuum chamber contains the pulse tube head, with vacuum ports providing access for temperature sensor and helium gas feedthroughs as well as for various vacuum connections and gauges. A radiation shield attached to the first stage of the pulse tube (at $\approx$ 30 K) reduces the heat load on the colder second stage (at $\approx 3$ K). Rectangular windows on both sides of the 30K shield allow optical access to the cell and along the beam line. A hole at the front of the 30K shield allows extraction of the molecular beam.

The cell is attached to a 3K cold plate bolted to the second stage of the pulse tube. A 3K shield reduces the blackbody heat load on the cell. Windowless holes in the shield allow optical access to the cell and along the beam line, while a hole in the front enables beam extraction. The inside of the shield is covered with coconut charcoal (PCB 12$\times$30 mesh Calgon Charcoal), which acts as a cryopump for helium gas \cite{Tobin1987,Sedgley1987}; the charcoal is affixed to the shield with epoxy (Arctic Silver Thermal Epoxy).

Room temperature helium gas ($99.999\%$ purity) flows into the cell through a series of stainless steel and copper tubes.  The flow rate $\mathcal{F}$ is monitored outside the vacuum chamber (using an MKS 246 Flowmeter).  The helium gas first thermalizes to 30 K and then to 3 K via copper bobbins on the two cryogenic stages. Thin-walled stainless steel tubes thermally isolate the bobbin stages from each other and from room temperature. High-purity helium is used to reduce the risk of clogging the flow tubes through condensation of impurities in the gas.  The cooled helium enters the back of the cell through a 3.2 mm OD copper tube. The cell is formed by drilling two perpendicular holes (22.9 mm diameter) into a copper block, giving the cell an interior volume of $\approx$ 15 cm$^3$ with characteristic size $L_c \approx 2.5$ cm. The size of the cell was chosen to be small such that large extraction of molecules into the beam could be achieved (see Eq. (\ref{eq:efficiency})).  The ablation target is mounted at 45 degrees relative to the molecular beam axis on a copper holder near the helium gas inlet.  Cell windows are uncoated sapphire for maximum thermal conductivity and are sealed to the cell with indium for good thermal contact. AR coated windows are avoided since they tend to become opaque, presumably from reacting with products of the ablation.

A mixture of helium, SrF, and other particles created via ablation exit the cell through a $d=3$ mm diameter hole in a 0.5 mm thick copper plate at the front of the cell to form a beam.  The beam passes through a 6 mm diameter hole in a coconut charcoal-covered 3K copper plate, typically placed 34 mm from the cell.  This plate acts to reduce the helium gas load into the rest of the apparatus.  73 and 86 mm from the cell, the beam passes through holes in the 3K and 30K shields respectively, and propagates into a room temperature vacuum apparatus.

\begin{figure}
\includegraphics[height=2.5in]
{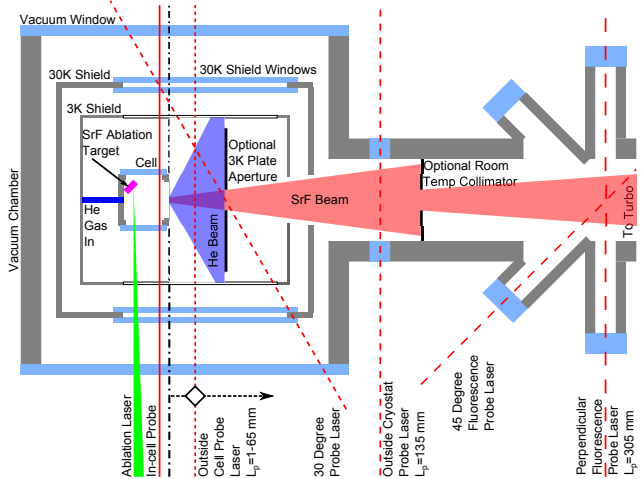} \caption{Experimental setup. Absorption measurements were made in-cell, perpendicular to the beam with $1$ mm $ < L_p <$ 65 mm, at 30 degrees to the beam at $L_p=20$ mm, and perpendicular to the beam outside the cryostat at $L_p =$ 135 mm. Fluorescence measurements were made perpendicular to and at 45 degrees to the beam at $L_p =305$ mm using the octagonal room-temperature apparatus.}  \label{fig:experimentdiagram}
\end{figure}

We optically probe the SrF molecules at various distances $L_p$ from the exit aperture to measure the characteristics of the molecular beam. We have optical access to the molecular beam for  1 mm $< L_p<$ 65 mm through rectangular holes in the vacuum chamber and the radiation shields.   All measurements for $L_p >$ 65 mm occur in the room-temperature beam region.  This region consists of either a small cross or an octagonal structure.  The cross has two viewports perpendicular to the molecular beam, and allows optical access to the beam at $L_p =135$ mm.  The octagon has 6 viewports, two perpendicular to the molecular beam and four oriented at $\pm 45^\circ$, and allows measurements at $L_p =  305$ mm. Fig. \ref{fig:experimentdiagram} depicts the apparatus.

A Nd:YAG laser (Big Sky Laser CFR200) produces 25 mJ pulses of 1064 nm light with $\approx 10$ ns pulse duration for ablation of the molecular precursor target. The laser beam is expanded through a telescope to a diameter of $\sim$ 15 mm before being tightly focused onto the ablation target by a lens of focal length $f = 20$ cm.  We note that the optimal conditions for focusing are observed to differ between species and targets; this configuration represents the optimum for production of SrF from our SrF$_2$ targets. Ablation targets are typically made by subjecting anhydrous SrF$_2$ powder (Sigma Aldrich 450030) to a pressure of 600 MPa using a die (Carver 3619) and a hydraulic press.

The continuous operation time of the beam is limited in part by saturation of the charcoal cryopump. We use $\approx$ 400 cm$^2$ of charcoal and find that it is adequate to allow run times of $>$ 20 hours at $\mathcal{F}=5$ sccm. At this flow rate we estimate the vacuum to be $\sim 4 \times 10^{-8}$ Torr inside the cryogenic region based on measured pumping speeds for coconut charcoal \cite{Tobin1987,Sedgley1987}. Once the charcoal is saturated, it must be warmed to $\gtrsim$ 20 K to allow the He to desorb and then be removed by the room-temperature vacuum pumps. The regeneration process for the charcoal cryopump (including the subsequent cooldown) takes $\sim$ 1 hour. Outside the cryogenic region, a 70 L/s turbo pump maintains the vacuum at $\sim 5 \times 10^{-7}$ Torr for a flow rate of $\mathcal{F}=5$ sccm.

\begin{figure}
\includegraphics[height=2.6in]
{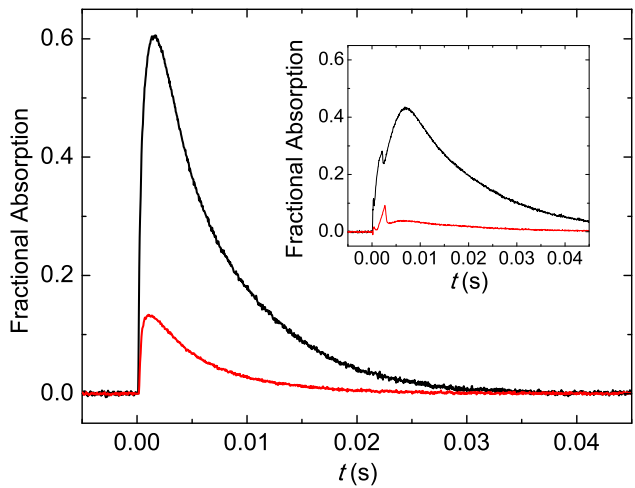} \caption{Typical raw absorption traces in the cell ({\color{black}\textbf{------}}) and immediately outside at $L_p = 1$ mm ({\color{red}\textbf{------}}) for $\mathcal{F} =  5$ sccm (main figure) and $\mathcal{F}=$ 50 sccm (inset). The in-cell and beam time traces are very similar to each other for each flow rate.  For short times a rapid increase in the absorption signal occurs as population in the X ($N_{rot}=0$) state increases via thermalization and passes through the probe laser.  At long times the absorption signal decreases as molecules are lost from the cell through the exit aperture, and through collisions with the cell walls.} \label{fig:typicalabsorptiontrace}
\end{figure}

\begin{figure}
\includegraphics[height=2.6in]
{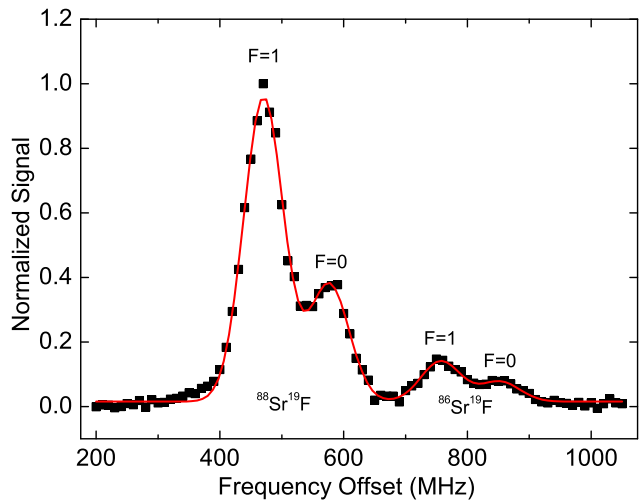} \caption{Typical in-cell absorption Doppler profile.  This spectrum was recorded by scanning a probe laser located inside the cell over the X ($v=0,N_{rot}=0$) $\rightarrow$ A ($v'=0,J'=3/2^-$) transition.  This plot shows the total time-integrated absorption (traces similar to those shown in Fig. \ref{fig:typicalabsorptiontrace}) for each laser frequency.  The four peaks correspond to the two $F=0,1$ hyperfine levels for the two $^{88}$Sr$^{19}$F and $^{86}$Sr$^{19}$F isotopes as labeled.  The hyperfine levels exhibit the expected 3:1 signal height ratios due to their Zeeman degeneracy, while the isotopes exhibit the expected 8.5:1 height ratio based on their natural abundance.  The solid line indicates the Doppler fit as described in the main text.}  \label{fig:integratedabsorptionlineshape}
\end{figure}

\section{Experimental Results}
We probe the number, temperature, and velocities of SrF in the cell and in the beam with resonant laser light from an external cavity diode laser using either absorption or laser-induced fluorescence (LIF).   Unless explicitly noted otherwise, the laser drives $Q_{11}(1/2)$ transitions from the ground state, X$^2\Sigma^+_{1/2}(v=0,N_{rot}=0,J=1/2)$, to the electronically excited state, A$^2\Pi_{1/2}(v'=0,J'=1/2^-)$, at 663 nm as defined in \cite{Shuman2009}.

\subsection{In-cell Dynamics and SrF Properties}
\subsubsection{Thermalization}
In Fig. \ref{fig:typicalabsorptiontrace} we show typical in-cell absorption time traces.  At time $t=0$ the ablation laser pulse fires.  For the first $\sim 500$ $\mu$s after the ablation pulse, the number of molecules in the $N_{rot}=0$ state rapidly increases as SrF molecules thermalize with the 3K helium gas and pass through the probe laser.  This process causes the initial sharp increase in the absorption shown in Fig. \ref{fig:typicalabsorptiontrace}.   Although we did not study in-cell thermalization in detail, we can compare this thermalization time to that predicted by Eq. (\ref{eq:thermalizationtime}).  There are no measurements of $\sigma_{SrF-He}$, so we assume that $\sigma_{SrF-He} \approx \sigma_{He-He} = 1.05 \times 10^{-14}$ cm$^2$ \cite{Dondi1969}.  The calculated value of $\tau_{Th}$ also depends on the value of $n_{He}$.  Here and for the remainder of this paper, $\kappa$ is estimated to be the geometric mean of the completely effusive and completely supersonic limiting cases, which yields $\kappa = 1.5$.  For  $\mathcal{F} =5$ sccm this corresponds to a density $n_{He}=3.5 \times 10^{15}$ cm$^{-3}$.  We also estimate here and throughout that the initial translational temperature of the SrF molecules is $T_{SrF}(0) = 10^4$ K \cite{Davis1985}. However we note the thermalization characteristics of this model depend only weakly on $T_{SrF}(0)$. Under these assumptions we obtain $\tau_{Th}\approx250$ $\mu$s, which is in reasonable agreement with our observations.

After the initial thermalization time, the absorption signal peaks and then decays as the molecules diffuse throughout the cell to the walls, and are pumped out the exit aperture.  By fitting this decay to an exponential, we can determine the molecule removal time constant $\tau_{rmvl}$.  For example, we find $\tau_{rmvl}= 7$ ms for the in-cell data with $\mathcal{F}=5$ sccm shown in Fig. \ref{fig:typicalabsorptiontrace}.   We find that $\tau_{rmvl}$ depends critically on ablation parameters.  Under only nominally different ablation locations on the target, or slightly different YAG focusing conditions, we observe that $\tau_{rmvl}$ can vary by a factor of 2 or more.  This indicates that the simple model of diffusion and extraction is heavily perturbed by the ablation process.  We also find that for high flow rates, $\mathcal{F}\gtrsim 30$ sccm, the temporal shape of the absorption signal changes significantly as shown in the inset of Fig. \ref{fig:typicalabsorptiontrace}, indicating more complicated in-cell processes than just simple diffusion.  Nonetheless, by fitting the decay at long times to an exponential, we find that $\tau_{rmvl}$ is of the same order of magnitude as either $\tau_{pump}^{He}$ or $\tau_{diff}$ for all flow rates investigated. For example, at $\mathcal{F}=5$ sccm, $\tau_{diff}=1.2$ ms, $\tau_{rmvl}=7$ ms, and $\tau_{pump}^{He}=25$ ms.

\subsubsection{In-cell Translational, Rotational and Vibrational Temperature}
During thermalization, collisions with $b$ cause the in-cell translational temperature of SrF, $T_{SrF}^{cell}$, to cool. We obtain $T_{SrF}^{cell}$ and in-cell velocity distributions for the molecules by incrementally scanning the probe laser frequency and recording a signal trace in time for each discrete frequency. The raw signals are integrated in time, starting 300 $\mu$s after the ablation for a duration of 20 ms unless explicitly noted otherwise. A typical in-cell absorption spectrum is shown in Fig. \ref{fig:integratedabsorptionlineshape}.  The four peaks correspond to the two $F=0,1$ hyperfine levels for $^{86}$Sr$^{19}$F and $^{88}$Sr$^{19}$F.  The integrated signal versus frequency lineshape is then fit to a sum of four Gaussians. The relative amplitudes of the Gaussians are constrained by the known abundance of the Sr isotopes and the Zeeman degeneracies for the hyperfine levels.  The widths of the Gaussians are constrained to be the same for each peak.  From the fitted width we extract a translational Doppler temperature and a velocity distribution.

We find $T^{cell}_{SrF}\approx5$ K over the full range of flow rates investigated. The value of $T^{cell}_{SrF}\approx5$ K we observe is larger than the temperature of the cell, $T_0 \approx 3$ K.  This is believed to be due to the initial heating of the buffer gas by the ablation of the target.  In support of this claim, we have measured the in-cell translational Doppler width in 1 ms time increments after ablation and found that the translational width decreases at longer times ($t \gtrsim 2$ ms), despite the fact that these observation times are very long compared to $\tau_{Th}$. This type of behavior has also been observed in other similar experiments \cite{Skoff2010}.

Thermalization also cools the in-cell rotational temperature, $T_{rot}^{cell}$. For temperatures $\sim 4$ K, typically $\sigma_{SrF-He}/\sigma^{rot}_{SrF-He} \sim 10-100$ \cite{Ball1999,Krems2009}. However, due to the large mass mismatch between SrF and He, overall we expect both translational and rotational thermalization to occur with similar efficiency.
We determine $T_{rot}^{cell}$ by comparing the relative populations in the X ($N_{rot}=0-4$) states using the X$^2\Sigma^+_{1/2}$($v=0,N_{rot}=0-4$) $\rightarrow$ A$^2\Pi_{1/2}$($v'=0,J'=1/2^{-}-9/2^{-}$) transitions.  We then fit the relative populations to a Boltzmann distribution.  As shown in Fig. \ref{fig:rotationalinsideoutsidechecked}, inside the cell we find $T_{rot}^{cell}=5.3$ K, comparable to results obtained with a similar apparatus and another molecular species \cite{Lu2009}.  We note $T^{cell}_{SrF}\approx T_{rot}^{cell}$ as expected.

Vibrational temperatures are expected to thermalize much more slowly than the rotational and translational temperatures because $\sigma^{vib}_{SrF-He}\ll\sigma_{SrF-He}$.   In Fig. \ref{fig:vibrationalpopulationschecked} we plot the relative populations of the first four vibrational levels $(v=0,1,2,3)$ inside the cell obtained by probing the X$^2\Sigma^+_{1/2}$($v=0-3,N_{rot}=0$) $\rightarrow$ A$^2\Pi_{1/2}$($v'=0-3,J'=1/2^{-}$) transitions.  As shown in Fig. \ref{fig:vibrationalpopulationschecked}, the data cannot be described by a Boltzmann distribution.  Nonetheless, we can roughly characterize the distribution by fitting the relative populations of the first two vibrational levels  to a Boltzmann distribution to yield the in-cell vibrational temperature, $T^{cell}_{vib} \sim\!300$ K.  We find that $T^{cell}_{vib}\gg T_{0}$ which indicates that the vibrational degree of freedom has not completely thermalized with the helium buffer gas.  Nevertheless, $T^{cell}_{vib}$ is still far lower than would be expected for that of unthermalized ablation products.

\begin{figure}
\includegraphics[height=2.6in]
{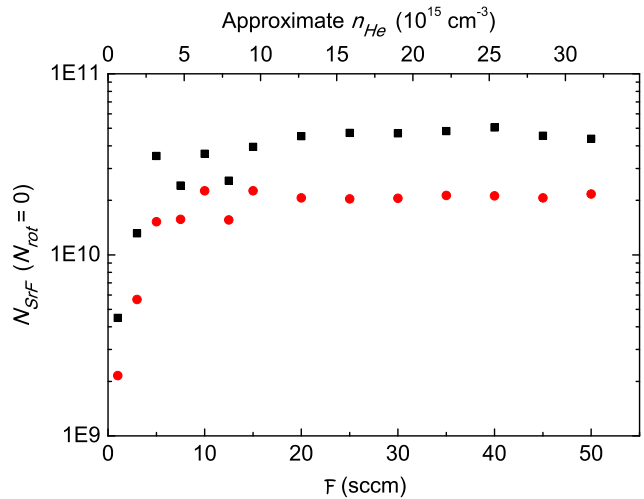} \caption{Number of SrF molecules $N_{SrF}^{cell}$ at $L_p=-1$ mm \small(\footnotesize${\blacksquare}$\small) and $N_{SrF}^{beam}$ at $L_p = +1$ mm \small(\large{\color{red}${\bullet}$}\small), as a function of $\mathcal{F}$. Although the parameter $\xi$ in Eq. (\ref{eq:efficiency}) varies between 0.01 and 0.5 over the flow rates examined here, the data indicate an extraction efficiency of $\epsilon \approx 0.5$ independent of $\mathcal{F}$.  } \label{fig:yieldversusflowrate}
\end{figure}

\subsubsection{SrF Yield}

In Fig. \ref{fig:yieldversusflowrate} we show the number of molecules in the $N_{rot} = 0$ state created in the cell, $N^{cell}_{SrF}$, for various flow rates and approximate values of $n_{He}$. The number of molecules is determined through the direct absorption of an in-cell probe laser with a diameter of 2 mm located at $L_p = -1$ mm (1 mm before the exit aperture).  For a resonant probe laser sampling species $a$ over a path length $L_s$, the ratio of the transmitted power $P_T$ to the initial power $P_0$ will vary as
\begin{equation}\label{eq:simpleabsorption}
\frac{P_T}{P_0} = e^{-n_a L_s \sigma_D}.
\end{equation}
Here $\sigma_D$ is the Doppler broadened absorption cross section \cite{Budker2008}, which is calculated from the lifetime of the A$^2\Pi_{1/2}$ state and the H\"{o}nl-London factors \cite{Sauer2009} for SrF. We use the peak absorption signal to calculate the in-cell number.  Using Eq. (\ref{eq:simpleabsorption}) and assuming a uniform $n_{SrF}$ within the entire volume of the cell, we obtain $N_{SrF}^{cell}$.

As shown in Fig. \ref{fig:yieldversusflowrate}, $N^{cell}_{SrF}$ increases with $\mathcal{F}$ for flows up to $F\sim\!5$ sccm, then reaches a maximum value of $N_{SrF}^{cell}\approx 4\times10^{10}$ and remains constant for higher flow rates.  We attribute the decrease in $N_{SrF}^{cell}$ at low flow rates to insufficient helium density to completely thermalize all the molecules.  The helium density at $\mathcal{F}=5$ sccm is $n_{He} \approx 3.5 \times 10^{15}$ cm$^{-3}$, which is in reasonable agreement with the minimum density required for thermalization as predicted by Eq. (\ref{eq:thermalizationdensity}), $n_{Th} \approx 1.5\times 10^{-15}$ cm$^{-3}$.

\subsection{Molecular Beam Properties}

In Fig. \ref{fig:typicalabsorptiontrace} we show a typical absorption time trace taken for $\mathcal{F}=5$ sccm with the probe laser located at $L_p = 1$ mm  (just outside the cell exit aperture).  The molecules that exit the cell exhibit a similar temporal profile as molecules in the cell.  As the SrF molecules exit the cell, the number of collisions between SrF and helium in and around the exit aperture largely determines the properties of the molecular beam far downstream.  We expect $Re_{SrF}>Re_{He}$ as discussed previously, but since we do not have an accurate value for $\sigma_{SrF-He}$, we solely use $Re_{He}$ to provide a qualitative indicator of whether the molecular beam should exhibit supersonic or effusive characteristics. Values of $Re_{He}$ are estimated using Eqns. (\ref{eq:meanfreepath}) and (\ref{eq:n}).

\subsubsection{Extraction from Cell}

We determine the number of molecules in the X ($N_{rot}=0$) state which exit the cell by measuring the absorption of a resonant probe laser with diameter small compared to $d$ and located at $L_p = 1$ mm. We then time-integrate the resonant absorption traces (similar to the one shown in Fig. {\ref{fig:typicalabsorptiontrace}). We also assume a uniform $n_{SrF}$ over the same area as the exit aperture and the measured Doppler spread ($\approx 5$ K).  The number of molecules in the beam at distance $L_p$ from the aperture, $N_a^{beam}$, can then be found using
\begin{equation}\label{eq:beamnumber}
N_a^{beam} = \frac{A_d v_{a\parallel}}{L_s \sigma_{D}}\int \ln \bigg[\frac{P_0}{P_T}\bigg]dt,
\end{equation}
where $P_0/P_T$ is the ratio of incident to transmitted power of the probe laser and $A_d$ is the cross sectional area of the molecular beam at $L_p$, determined either by geometric constraints after any collimators or by the measured divergence of the beam prior to any collimators. In Fig. \ref{fig:yieldversusflowrate} we plot $N_{SrF}^{cell}$ (at $L_p = -1$ mm) and $N_{SrF}^{beam}$ immediately outside the cell (at $L_p= 1$ mm) for various flow rates. By comparing the number of molecules inside and just outside the cell, we can determine the extraction efficiency $\epsilon$ for the cell.  Over the range of flows examined, the ratio of the estimated diffusion time to the estimated pumpout time, $\xi$, varies between 0.01 and 0.5.  Based on the extraction model presented earlier, we would crudely expect $\epsilon$ to vary between $\epsilon_{eff}\sim 0.003$ and $\epsilon_{hyd} \sim 1$ over this range.  Instead, we find that $\epsilon \sim\! 0.5$, independent of $\mathcal{F}$ over this range.  This suggests that the extraction model presented earlier is too simplistic to fully capture the dynamics inside the cell.

\begin{figure}
\includegraphics[height=2.6in]
{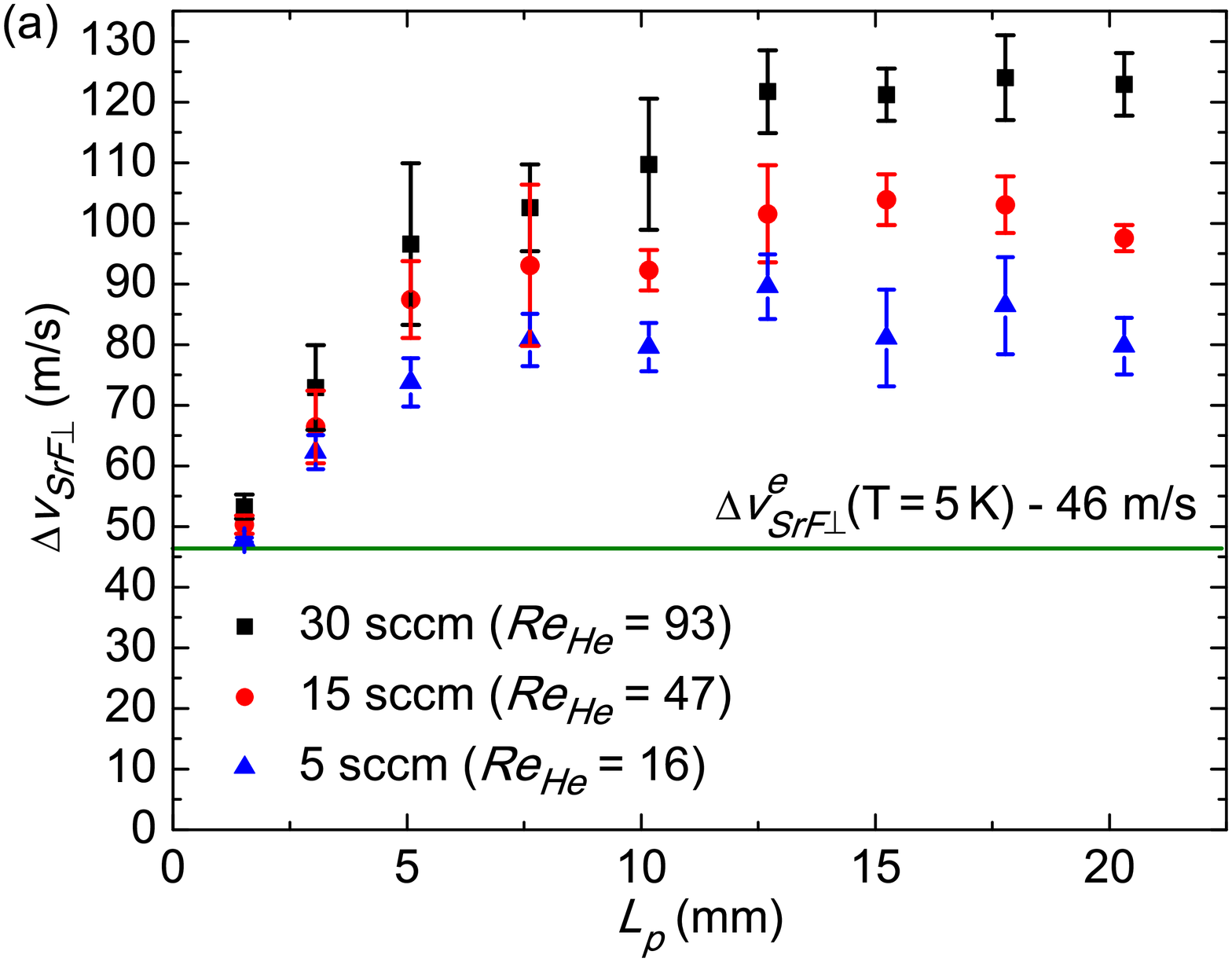}
\includegraphics[height=2.6in]
{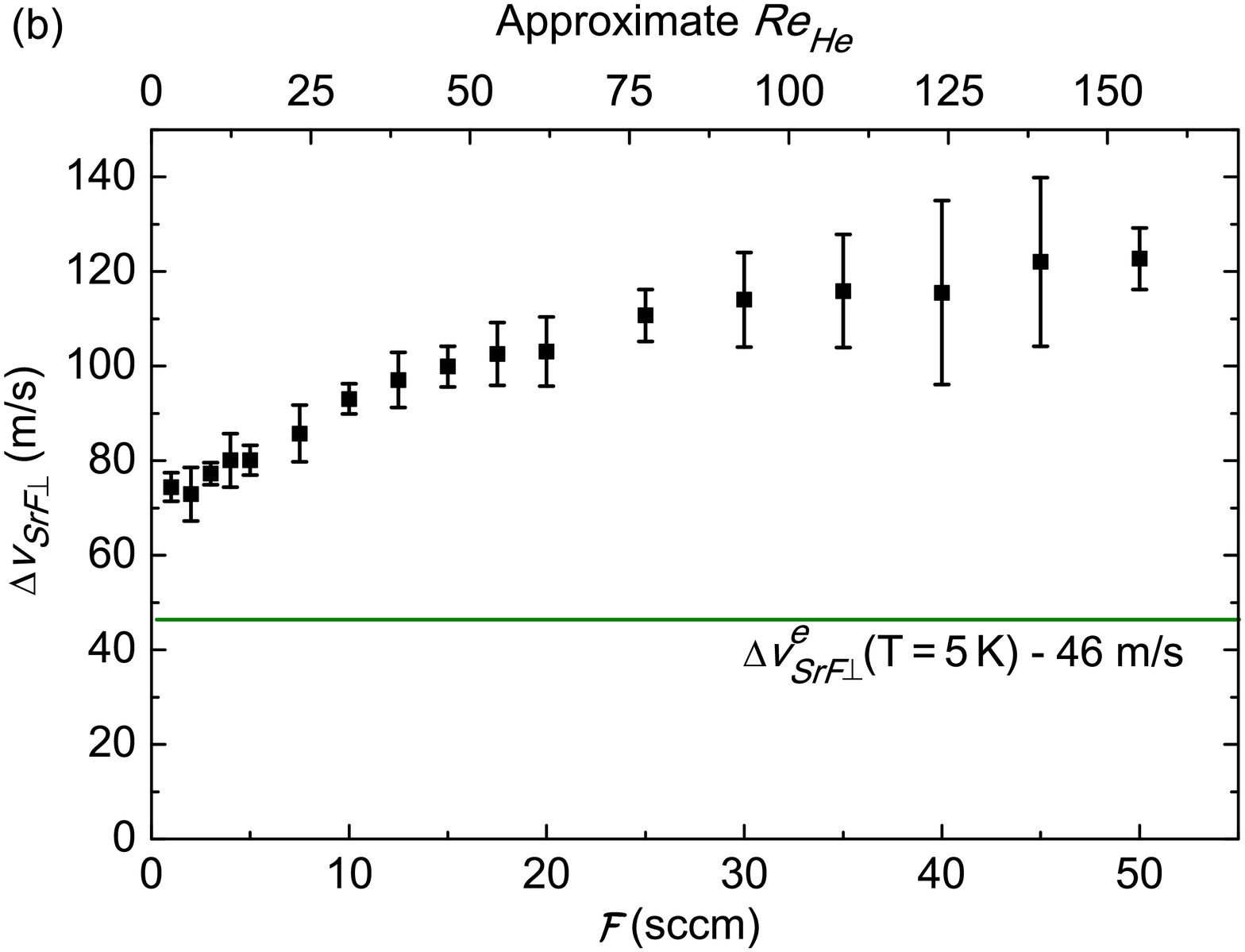}
\caption{(a) FWHM transverse velocity of the molecular beam  $\Delta v_{SrF\perp}$ versus probe distances $L_p$, for $\mathcal{F}=$ 5({\color{blue}$\blacktriangle$}), 15\small(\large{\color{red}$\bullet$}\small), and 30\small(\footnotesize${\blacksquare}$\small) sccm respectively. We observe $\Delta v_{SrF\perp}$ is consistent with a $\sim\!5$ K Boltzmann distribution for SrF just outside the cell, and that $\Delta v_{SrF\perp}$ increases with increasing probe distance $L_p$ before leveling off for $L_p\gtrsim 10$ mm. Larger values of $\mathcal{F}$ result in greater final values of $\Delta v_{SrF\perp}$. The value of $\Delta v_{SrF\perp}^{e}$ is calculated from \cite{Ramsey1956}.   (b) FWHM transverse velocity of the molecular beam versus $\mathcal{F}$ for $L_p=20$ mm, where the width is no longer increasing with distance from the cell.  We attribute the increase in $\Delta v_{SrF\perp}$ with increasing $\mathcal{F}$ to a helium pressure gradient outside the cell, as discussed in the main text. Error bars in this and in all figures hereon represent the standard deviation of a set of several (typically 3-10) data points taken under nominally identical conditions.} \label{fig:transversewidthversusdistancefromaperturechecked}
\end{figure}

\begin{figure}
\includegraphics[height=2.6in]
{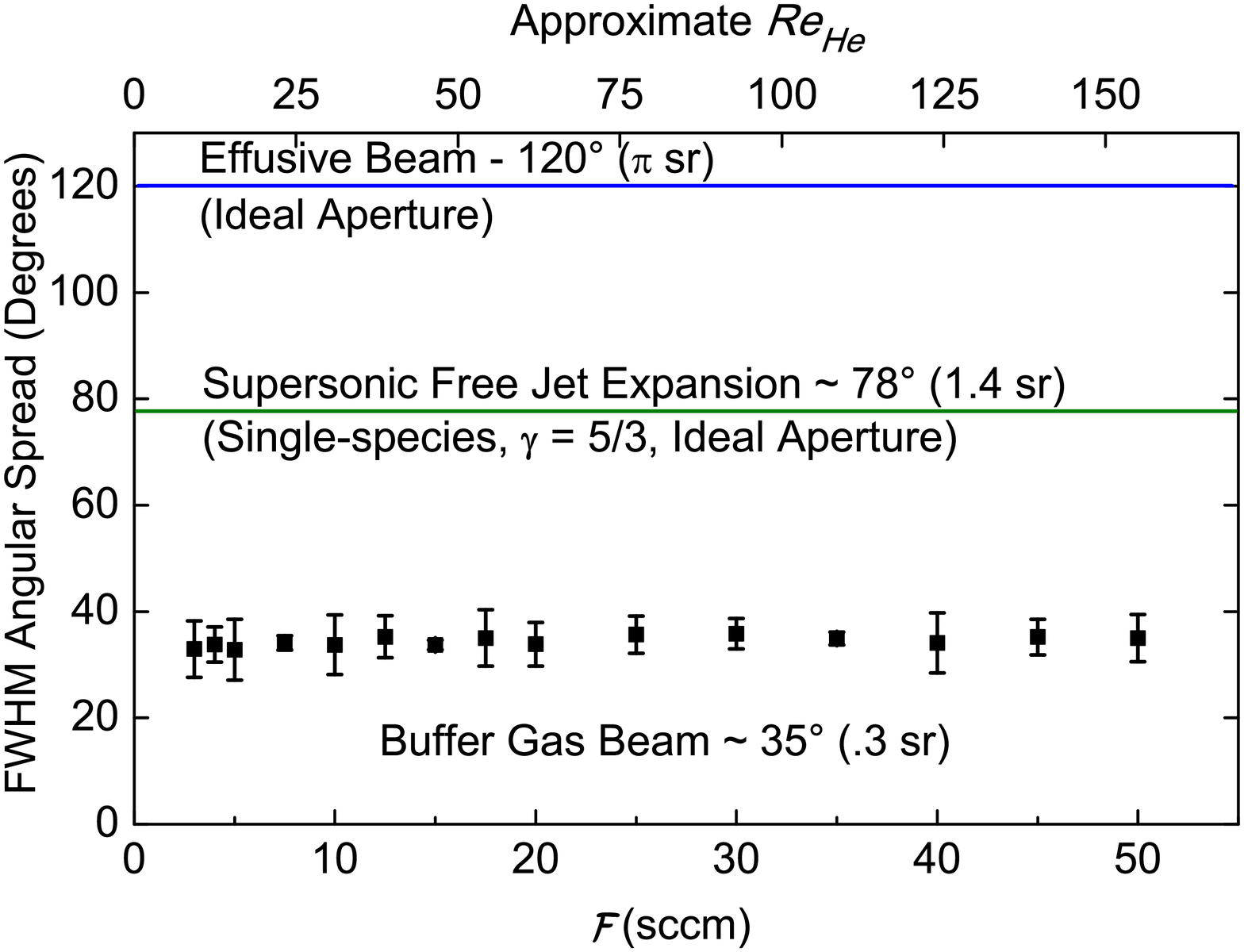} \caption{Molecular beam divergence for various flow rates. The final beam divergence demonstrates little dependence on $\mathcal{F}$.  Included is a comparison of beam divergences for different types of beams. The beam in this work is more directional than either an effusive beam or a single-species supersonic free jet expansion from an ideal aperture. The divergence of an effusive beam is calculated from \cite{Ramsey1956} while the divergence of a single-species supersonic free jet expansion from an ideal aperture is calculated from \cite{Tejeda1996}, in agreement with measurements \cite{Hillenkamp2003}.} \label{fig:beamdivergenceversusflowrate}
\end{figure}

\subsubsection{Beam Transverse Measurements}

As the SrF molecules propagate from the cell, they experience fewer collisions with the helium buffer gas, as its density falls as $1/L_p^2$.  However, the expansion of the helium gas into the vacuum outside the cell and ongoing helium-SrF collisions dramatically change the characteristics of the beam.  An example of this behavior is shown in Fig. \ref{fig:transversewidthversusdistancefromaperturechecked}(a).  Here we plot the FWHM transverse velocity spread,  $\Delta v_{SrF\perp}$, of the molecular beam as a function of $L_p$ for a few different values of $\mathcal{F}$.  Within one hole radius of the aperture, $\Delta v_{SrF\perp}$ was measured to be consistent with a $\sim\!5$ K Boltzmann distribution for SrF, independent of $\mathcal{F}$. This value is very similar to $T^{cell}_{SrF}\approx5$ K.  However, $\Delta v_{SrF\perp}$ increases at further distances downstream before leveling off to a constant value, as shown in Fig. \ref{fig:transversewidthversusdistancefromaperturechecked}(a), with larger $\mathcal{F}$ resulting in larger final values of $\Delta v_{SrF\perp}$. The broadening of $\Delta v_{SrF\perp}$ outside the aperture nozzle is in qualitative agreement with the presence of a He pressure gradient transverse to the molecular beam outside the cell \cite{Sherman1965,Dun1979,Ramos2009}.  This would cause the greatest rate of broadening closest to the cell aperture where pressure gradients are strongest. Larger values of $\mathcal{F}$ would also produce larger pressure gradients, resulting in greater broadening of $\Delta v_{SrF\perp}$. Additional broadening beyond $L_p\gtrsim 10$ mm is not observed, as shown in Fig. \ref{fig:transversewidthversusdistancefromaperturechecked}(a), indicating that collisions with helium no longer affect the characteristics of the SrF beam beyond this distance. Fig. \ref{fig:transversewidthversusdistancefromaperturechecked}(b) depicts $\Delta v_{SrF\perp}$ for $L_p=20$ mm for a variety of flow rates $\mathcal{F}$. Combining this data with the beam forward velocities measured in the next section, we determine the beam divergence to be nominally independent of $\mathcal{F}$ as shown in Fig. \ref{fig:beamdivergenceversusflowrate}. Compared to both an effusive beam and a single-species supersonic free jet expansion beam from an ideal aperture, the beam in this work is significantly more directional.

\begin{figure}
\includegraphics[height=2.6in]
{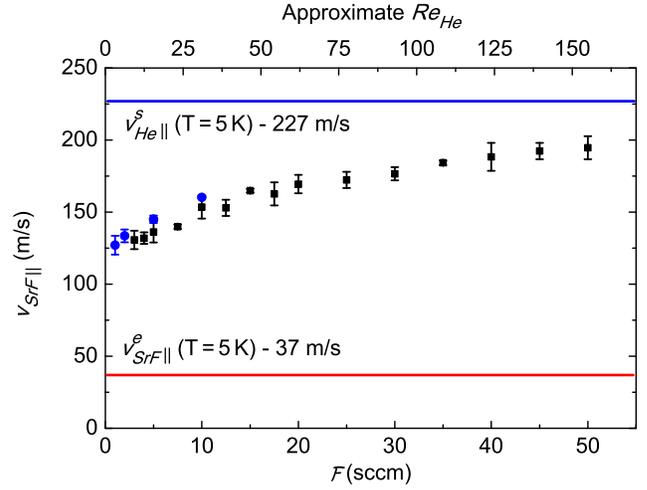} \caption{Forward velocity for various flow rates.  The data were taken at $L_p=15$ mm in absorption \small(\footnotesize${\blacksquare}$\small) and at $L_p =305$ mm using LIF \small(\large{\color{blue}$\bullet$}\small). For all $\mathcal{F}$ we observe $v_{SrF\parallel} > v_{SrF\parallel}^{e}$. Although we estimate $Re_{He}\sim1$ for the lowest $\mathcal{F}$, this observation indicates that there are still enough collisions in the aperture to boost the forward velocity of SrF above $v_{SrF\parallel}^{e}$.  At the highest $\mathcal{F}$ where $Re_{He}\gg1$, we observe $v_{SrF\parallel} <v_{He\parallel}^{s}$, as we expect.  Measurements at different values of $L_p$ are in good agreement.} \label{fig:forwardvelocityversusflowratechecked}
\end{figure}

\begin{figure}
\includegraphics[height=2.6in]
{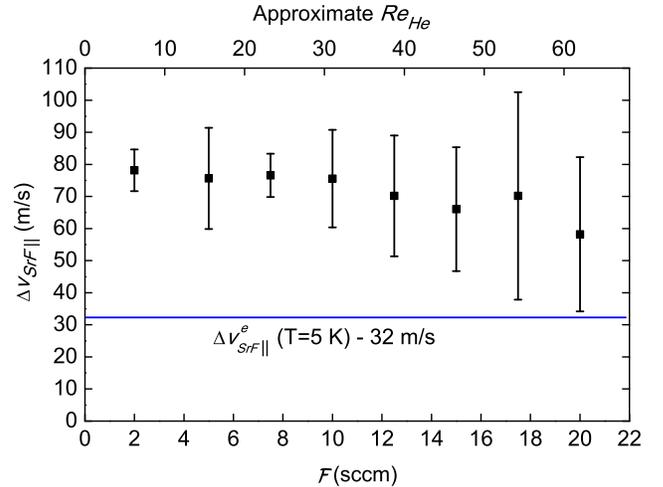} \caption{FWHM forward velocity spread for various flow rates.  The typical measured value of $\Delta v_{SrF\parallel} \approx 75 \frac{m}{s}$ is roughly constant across the range of $\mathcal{F}$ explored.  The value of  $\Delta v_{SrF\parallel}$ is well above the maximum expected value of $\Delta v_{SrF\parallel}^{e}$ ({\color{blue}\textbf{------}}), calculated from \cite{Ramsey1956}.  We note as well that for fixed $\mathcal{F}$, the measured values of $\Delta v_{SrF\parallel}$ varied appreciably ($\sim 40\%$) under nominally identical conditions.  These observations are not compatible with the simple thermalization model presented in the text, and suggest that the ablation significantly perturbs the in-cell thermalization process.} \label{fig:forwardvelocityspreadversusflowratechecked}
\end{figure}

\begin{figure}
\includegraphics[height=2.6in]
{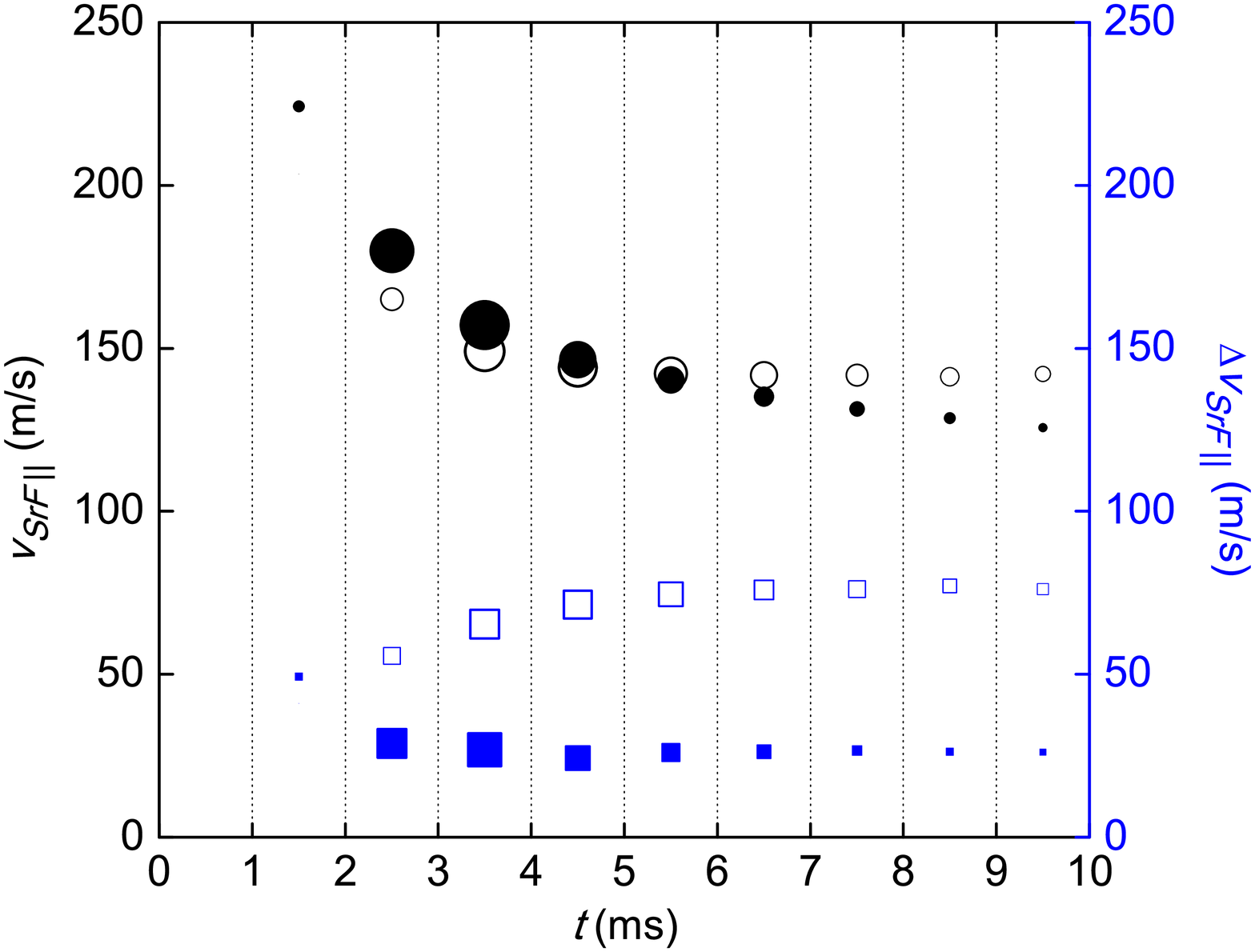} \caption{Measured forward velocity \small(\large{\color{black}$\bullet$}\small), measured forward velocity spread (\footnotesize{\color{blue}$\blacksquare$}\small), simulated forward velocity \small(\normalsize{\color{black}$\circ$}\small) and simulated forward velocity spread \small(\footnotesize{\color{blue}$\Box$}\small) versus time. The data were taken at $L_p=305$ mm in LIF using 1 ms time bins for a flow rate of $\mathcal{F}=5$ sccm. The size of the points indicate relative molecule number. The measured data indicate that both $v_{SrF\parallel}$ and $\Delta v_{SrF\parallel}$ vary significantly over the time the beam persists; molecules detected earliest have the largest $v_{SrF\parallel}$ and $\Delta v_{SrF\parallel}$, and both decrease with time.  The simulated data assume a time-invariant gaussian velocity distribution at the exit aperture with $v_{SrF\parallel}=145$ m/s and $\Delta v_{SrF\parallel}=75$ m/s; thus, the time dependence of the simulated data downstream is solely due to time of flight effects.  Although the simulated data exhibit similar time variation in $v_{SrF\parallel}$, the magnitude of the effect is not sufficient to completely describe the measurements.  Furthermore the simulation is in qualitative disagreement with the measured time variation in $\Delta v_{SrF\parallel}$.  These observations are consistent with initial heating of the buffer gas at early times, as described previously.}
\label{fig:binnedforwarddata}
\end{figure}

\subsubsection{Beam Forward Velocity and Temperature}
The remainder of the measurements are made for $L_p>10$ mm, where collisions within the beam have largely ceased, and the characteristics of the beam are expected to be static.  In Fig. \ref{fig:forwardvelocityversusflowratechecked} we show the measured dependence of the beam forward velocity, $v_{SrF\parallel}$, on the helium flow rate $\mathcal{F}$ taken at two different places downstream from the cell. For $L_p = 15$ mm, we determine $v_{SrF\parallel}$ by comparing the Doppler shifts of direct absorption profiles of two probe lasers, one normal to the molecular beam and one at 30 degrees relative to normal. For $L_p=305$ mm, the same technique is employed but using LIF instead.

For all values of $\mathcal{F}$, we observe $v_{SrF\parallel} > v_{SrF\parallel}^e$, indicating that there are still sufficient collisions near the aperture to cause significant increase in  $v_{SrF\parallel}$ even at the lowest flow rates where $Re_{He} \sim 1$. At the highest $\mathcal{F}$ where $Re_{He}\gg1$, we expect that $v_{SrF\parallel}$ should approach $v^s_{He\parallel}$, in agreement with our observations.  However even at the highest $\mathcal{F}$, $v_{SrF\parallel} < v_{He\parallel}^{s}$.  Since we cannot measure the forward velocity of the helium in the beam, $v_{He\parallel}$, this observation may simply result from $v_{He\parallel}<v^s_{He\parallel}$ with $v_{SrF\parallel} = v_{He\parallel}$.  It may also be due to the phenomenon known as velocity slip, where the speed of the seeded species does not get fully boosted to the speed of the carrier \cite{Mazely1995,Abuaf1967,Pauly2000,Scoles88,Dea2009}.

In Fig. \ref{fig:forwardvelocityspreadversusflowratechecked} we show the FWHM of the forward velocity, $\Delta v_{SrF\parallel}$, for various values of $\mathcal{F}$.   $\Delta v_{SrF\parallel}$ was measured in LIF by varying the frequency of a $45^{\circ}$ probe laser 305 mm downstream and integrating over the entire duration of the molecular beam pulse.  An aperture collimates the molecular beam so that the transverse Doppler width is reduced to near the natural linewidth of the X-A probe transition ($\approx 7 $ MHz).  The Doppler broadening from the forward velocity of the beam is substantially larger than this ($\sim \! 100$ MHz); thus, fitting these distributions to a Gaussian enables extraction of the forward temperatures of the molecular beam.  For the range of $\mathcal{F}$ explored, the typical measured values of $\Delta v_{SrF\parallel} \approx 75 \frac{m}{s}$ (corresponding to $T_{SrF\parallel}\approx 13$ K) are well above $\Delta v_{SrF\parallel}^e$. This is in contrast to the behavior of typical seeded free jet expansions where $\Delta v_{SrF\parallel}<\Delta v_{SrF\parallel}^e$ due to cooling during the isentropic expansion.

A number of observations regarding the forward velocity suggest that the simplistic thermalization model described previously may not be adequate to describe this system.  For example, we find that the measured values of $v_{SrF\parallel}$ and $\Delta v_{SrF\parallel}$ vary by $\sim \! 15\%$ under nominally similar ablation conditions, depending on the location ablated on the target.  Specifically, ablating closest to the He gas inlet and furthest from the exit aperture tended to produce the lowest $v_{SrF\parallel}$ and $\Delta v_{SrF\parallel}$.  We also find that the characteristics of the ablation laser also change  $v_{SrF\parallel}$ and $\Delta v_{SrF\parallel}$.  In particular, the focus and the power of the ablation laser can alter $v_{SrF\parallel}$ and $\Delta v_{SrF\parallel}$ by  $\sim \!15\%$.  Finally, we observe very fast molecules ($\gtrsim 225$ m/s) with large $\Delta v_{SrF\parallel}$ at early times ($t\lesssim2$ ms) in the molecular pulse, as shown in Fig. \ref{fig:binnedforwarddata}.  The observed time variation in $v_{SrF\parallel}$ and $\Delta v_{SrF\parallel}$ cannot be fully explained by time of flight effects, which suggests that molecules leaving the cell at different times thermalize to different temperatures.  These observations are difficult to explain using the simple models outlined earlier, and suggest more complicated in-cell dynamics.

\begin{figure}
\includegraphics[height=2.6in]
{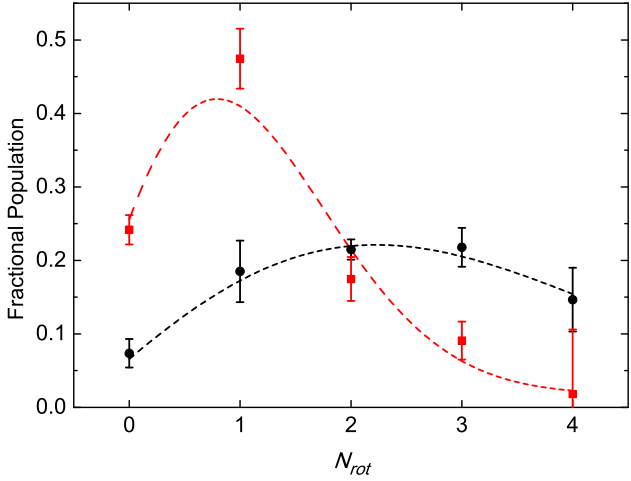} \caption{Fractional rotational populations in-cell \small(\large{\color{black}$\bullet$}\small) and at $L_p = 20$ mm downstream \small(\footnotesize{\color{red}$\blacksquare$}\small), with associated fits to a Boltzmann distribution. This data was taken at $\mathcal{F}=5$ sccm. The fits indicate $T^{cell}_{rot} = 5.3$ K and $T^{beam}_{rot}$($L_p=20$ mm) $=1.2$ K; this shows substantial rotational cooling as the beam leaves the cell.} \label{fig:rotationalinsideoutsidechecked}
\end{figure}

\begin{figure}
\includegraphics[height=2.6in]
{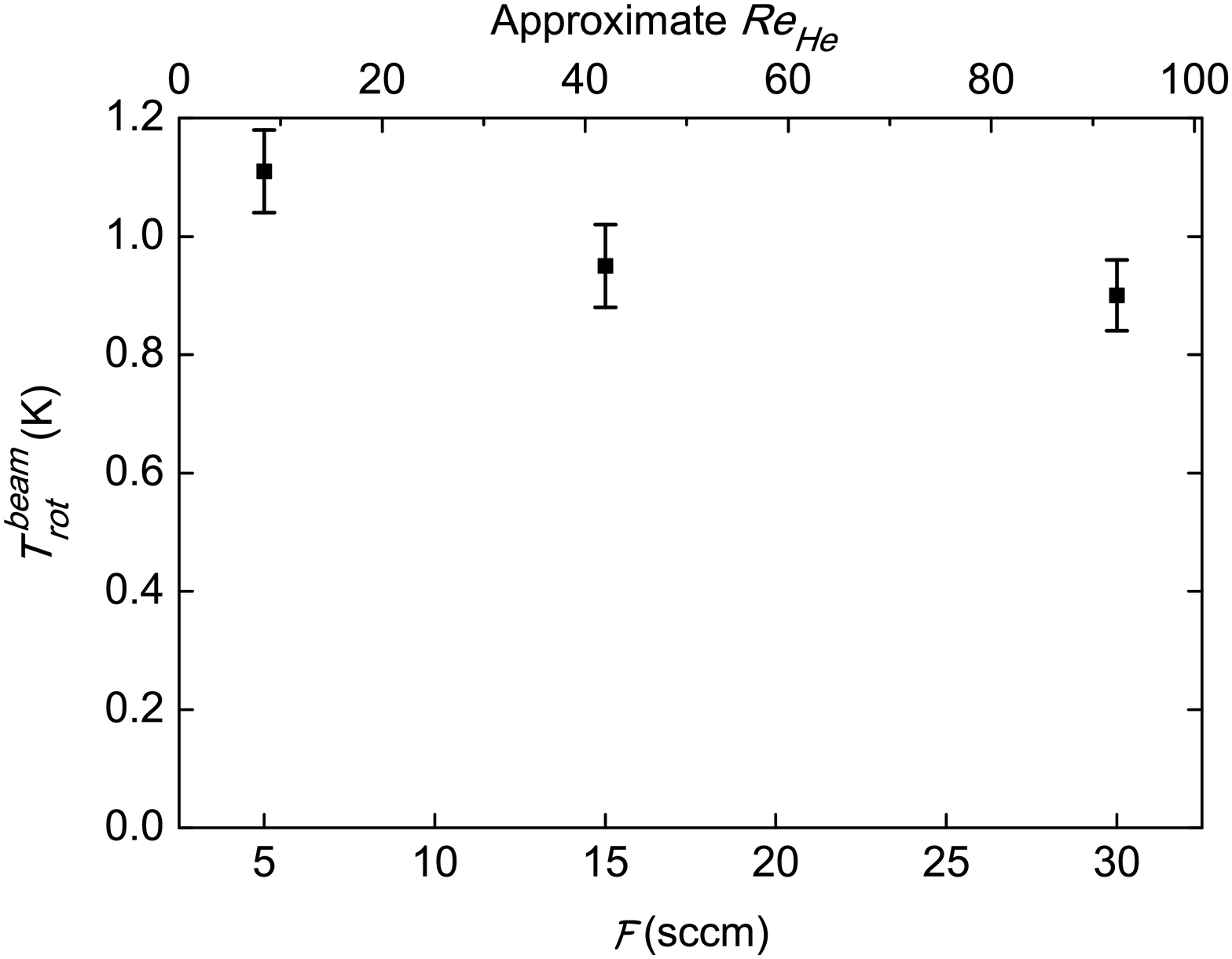} \caption{$T_{rot}^{beam}(L_p=20$ mm) versus $\mathcal{F}$. For the range of $\mathcal{F}$ investigated, rotational temperatures in the beam are $\approx$ 1 K. While we expect $T_{rot}^{beam}$ to decrease with increasing $\mathcal{F}$, these observations are consistent with other free jet sources that show a termination of rotational cooling.  These temperatures were determined by a fit to data similar to those shown in Fig. \ref{fig:rotationalinsideoutsidechecked}. } \label{fig:rotationaltemperatureversusflowratechecked}
\end{figure}

\begin{figure}
\includegraphics[height=2.6in]
{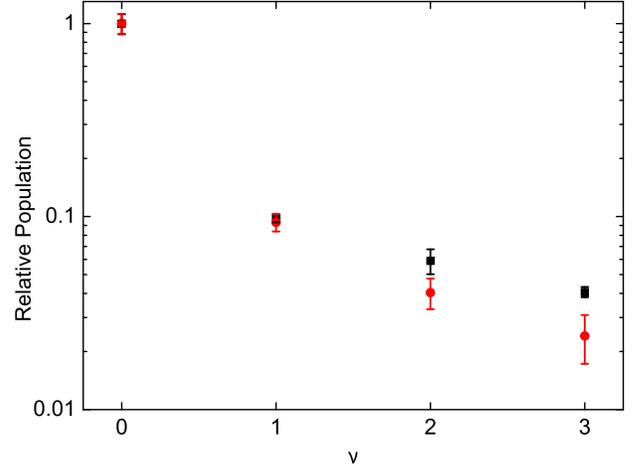} \caption{Relative vibrational populations in-cell \small(\footnotesize${\blacksquare}$\small) and at $L_p =$ 10 mm \small(\large{\color{red}$\bullet$}\small), at $\mathcal{F}=5$ sccm. Both datasets are normalized to 1 at $v = 0$.  The data cannot be fit to a simple Boltzmann distribution as the population does not decrease exponentially with $v$.  This indicates that the vibrational degree of freedom does not completely thermalize with the helium buffer gas.  If we extract a temperature from the ratio of populations in only the first two vibrational levels, we obtain $T^{cell}_{vib}\approx T^{beam}_{vib} \approx$ 300 K. This is substantially higher than $T_0$, but still significantly lower than the initial temperature of the SrF molecules created through ablation.  We observe little or no cooling of the vibrational degree of freedom in the beam.} \label{fig:vibrationalpopulationschecked}
\end{figure}

\subsubsection{Beam Rotational and Vibrational Temperature}

In Fig. \ref{fig:rotationalinsideoutsidechecked} we show the relative rotational populations measured at $L_p = 10$ mm for $\mathcal{F}=5$ sccm.  This distribution is consistent with a beam rotational temperature of $T_{rot}^{beam}=1.2$ K and is significantly colder than the measured in-cell rotational temperature of $T^{cell}_{rot}=5.3$ K. The observation of $T^{cell}_{rot} > T^{beam}_{rot}$ is attributed to cooling of rotational degrees of freedom via collisions near the aperture, as is typically observed in standard free jet expansions \cite{Scoles88,Pauly2000}.

In a separate measurement at $L_p =$ 20 mm downstream, similar rotational cooling was observed for a variety of $\mathcal{F}$, as shown in Fig. \ref{fig:rotationaltemperatureversusflowratechecked}.  We find that $T^{beam}_{rot}\approx1$ K for all flow rates investigated. Since the number of collisions outside the aperture is expected to scale linearly with $\mathcal{F}$, it is interesting that there is little change in $T^{beam}_{rot}$ over the range $\mathcal{F}$ = 5-30 sccm. The measured values of $T^{beam}_{rot}$ may be compared to the conservative upper limit on the ultimate downstream He temperature at these flow rates using Eq. (\ref{eq:supersonicfinaltemp}).  In this limit we find $T_{He}^{beam}$ $<$  2.38 K, 0.99 K and 0.57 K for $\mathcal{F}$ = 5, 15 and 30 sccm respectively. Similar results demonstrating $T_{rot}^{beam}$ largely independent of backing pressure have been observed for CO seeded in room-temperature He \cite{Ahern1999}.

In Fig. \ref{fig:vibrationalpopulationschecked} we plot the relative vibrational populations in the beam at $L_p$ = 10 mm.  Little if any cooling of the higher vibrational levels in the beam was observed; this is consistent with the notion that many more collisions are required to thermalize the vibrational degree of freedom \cite{Krems2009} than the rotational or translational degrees of freedom.

\begin{figure}
\includegraphics[height=2.6in]
{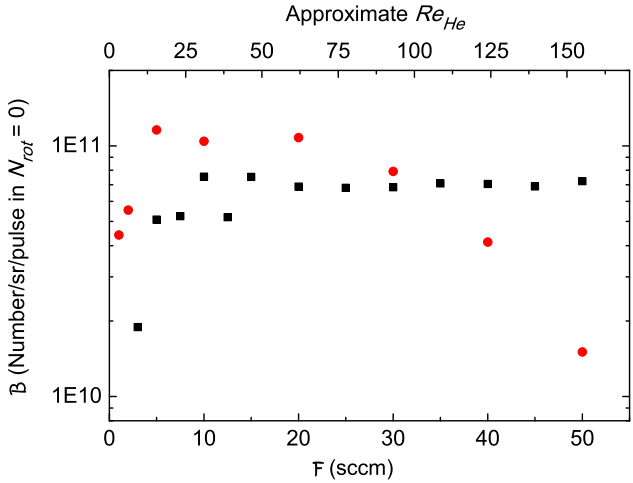} \caption{$\mathcal{B}$ for various flow rates $\mathcal{F}$ measured at $L_p =$ 1 mm \small(\footnotesize${\blacksquare}$\small) and $L_p =$ 135 mm \small(\large{\color{red}$\bullet$}\small). The data indicate the beam is attenuated over long distances, when the cell is operated at flow rates $F\gtrsim 20$ sccm. The peak downstream signal (at $\mathcal{F}=5-20$ sccm) corresponds to $\mathcal{B} \approx 1.2\times 10^{11}$ $N_{rot}=0$ molecules/sr/pulse. The increased $\mathcal{B}$ downstream versus upstream at low flow rates is likely due to rotational cooling in the beam. Downstream data were taken with the plate at $L_d=34$ mm.} \label{fig:brightnessversusflowrate}
\end{figure}

\subsubsection{Beam Brightness}

Finally, we have measured $\mathcal{B}$, the brightness of the molecular beam in the X ($N_{rot}=0$) state, both just outside the cell (at $L_p = 1$ mm) and at $L_p=135$ mm in a room-temperature environment, as shown in Fig. \ref{fig:brightnessversusflowrate}.   Inside the cryostat, geometrical constraints make further slowing, trapping, or precision spectroscopy of the molecular beam quite challenging, so $\mathcal{B}(L_p = 135$ mm) is representative of the useful beam brightness for most experiments.  To calculate $\mathcal{B}$ downstream we use Eq. (\ref{eq:beamnumber}).  A collimating aperture constrains the molecular beam to $\sim 0.03$ sr, so we assume a uniform $n_{SrF}$ over this solid angle and the Doppler spread associated with this geometry. Nominally, we expect $\mathcal{B}$ to remain constant as the beam propagates.  However, we observe $\mathcal{B}(L_p = 135 \text{ mm})/\mathcal{B}(L_p = 1 \text{ mm}) > 1$ for low $\mathcal{F}$.  We attribute this increase in observed downstream brightness to rotational cooling during the the first $\sim$ 10 mm of beam propagation.  For high $\mathcal{F}$, $\mathcal{B}(L_p = 135 \text{ mm})/\mathcal{B}(L_p = 1 \text{ mm}) < 1$, indicating a loss of molecules during beam propagation.  Because this loss increases with larger $\mathcal{F}$, we believe that the cause is a higher helium gas load, which can lead to a larger background density of helium and hence collisional attenuation of the SrF beam.

In our initial experiments, the ratio $\mathcal{B}(L_p = 135 \text{ mm})/\mathcal{B}(L_p = 1 \text{ mm})$ was significantly worse at high flow rates.  We found that placing a charcoal-covered plate (2.5 mm thick, with a 6.35 mm diameter hole) in the beam line substantially reduced this problem.  We suspect that the plate provides strong pumping of He gas near the beam axis, creating a differentially-pumped region behind the plate through which the beam can travel through without undergoing collisions with background helium. This plate was tested in two separate positions, at $L_d=21$ mm and $L_d=34$ mm downstream from the cell aperture; both placements largely eliminated beam brightness decreases for $F\lesssim20$ sccm. Ultimately we find that $L_d=34$ mm results in the highest brightness at $L_p=135$ mm for $\mathcal{F}\lesssim20$ sccm.  For  $\mathcal{F} \gtrsim 20$ sccm, we still observe a significant reduction in $\mathcal{B}$.  We did not investigate this further because we plan to primarily operate the apparatus in the low flow ($\mathcal{F}\lesssim20$ sccm) regime where the forward speeds are the lowest.  We speculate that the use of a true molecular beam skimmer might help alleviate this problem.

\begin{figure}
\includegraphics[height=2.6in]
{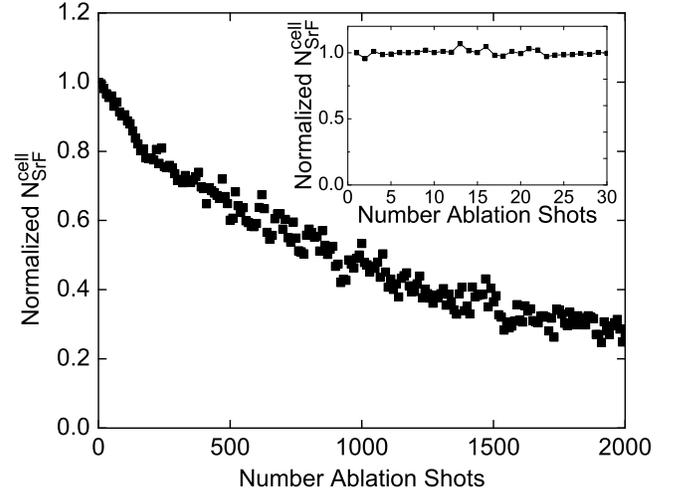} \caption{Normalized $N^{cell}_{SrF}$ (which is well-correlated with $N^{beam}_{SrF}$) as a function of ablation shot number. The same ablation spot was used for all 2000 shots to illustrate the durability of the target. $N^{cell}_{SrF}$ decays to half its initial value in 500-1000 shots. In the main figure, each data point represents the average of 10 ablation shots. The inset shows $N^{cell}_{SrF}$ for 30 consecutive shots to illustrate the shot-to-shot fluctuations in $N^{cell}_{SrF}$.  The variation shown here ($<10\%$) is typical for this system.  For the data shown in the inset, the pulse tube refrigerator was turned off, to distinguish ablation variation from the 1.4 Hz pulse tube temperature variation. } \label{fig:targetdurability}
\end{figure}

\subsubsection{Ablation, Noise and Variation}

We typically ablate the SrF$_2$ target at $R_{\text{YAG}} =1$ Hz rep. rate.  We observe that operation at $R_{\text{YAG}} > 1$ Hz sometimes decreases $N_{SrF}^{cell}$ by a factor of 2 or more. Other times $N_{SrF}^{cell}$ is nominally independent of $R_{\text{YAG}}$, up to $R_{\text{YAG}}\sim$ 15 Hz. This dependence on $R_{\text{YAG}}$ varies from target to target, and from spot to spot on the same target.

For consecutive shots on the same spot on the target,  $N_{SrF}^{cell}$ typically varies by $\sim5\%$ from shot to shot. Thus to produce consistent data, the ablation spot was changed only when necessary. Ablation yield from a single spot on the target was found to decrease after many shots. Typically $N_{SrF}^{cell}$ decreased by a factor of 2 after 500-1000 shots on the same spot (although this could vary by a factor of 2 or more). The steady decrease in $N_{SrF}^{cell}$ versus shot number, depicted in Fig. \ref{fig:targetdurability}, is typical.

The in-cell ablation yields are observed to vary significantly (a factor of 2 or more) for different nearby spots on the target. However, in the absence of significant visible damage to the window, an ablation spot can generally be found which will yield very nearly the maximum $N_{SrF}^{cell}$ from that target. Finding such optimal spots typically requires sampling of a dozen or so different ablation spots.

In an effort to improve yield, durability, or allow consistent operation at higher $R_{\text{YAG}}$, we investigated different ablation targets: a SrF$_2$ single crystal, a commercial isostatically hot-pressed SrF$_2$ target, and eight cold-pressed targets made in-house with the same procedure but using different precursor materials. The in-house targets used pure powders of anhydrous SrF$_2$, precipitated SrF$_2$, 1 $\mu$m SrF$_2$ and crushed macroscopic crystals of SrF$_2$, as well as the same powders mixed with powdered boron metal in a 1/9 molar ratio. While the yield from all targets was the same to within a factor of $\sim$ 2, ultimately the anhydrous SrF$_2$ with powdered boron metal offered the greatest yield and allowed rep. rates up to 15 Hz, equal to the best rep. rates of the group. All targets lasted for the same number of shots to within a factor of $\sim$ 2.

Ablation of SrF$_2$ produced macroscopic amounts of dust inside the cell. However, this dust did not create any known problems. After more than $10^6$ total ablation shots, both cell windows were visibly covered with dust (resulting in less than 10$\%$ transmission of a cw probe laser through each cell window), but $N_{SrF}^{cell}$ was not significantly affected. Thus $10^6$ can be taken as a lower bound on the number of ablation shots possible before the apparatus must be opened and the target replaced. The ablation laser appears to remove any dust from the window in its path.

In addition to variation due to ablation, the periodic temperature oscillation of the pulse tube refrigerator's second stage (1.4 Hz period, $T_{min}=2.85$ K, $T_{max}=3.15$ K) was observed to affect both $N_{SrF}^{cell}$ and $N_{SrF}^{beam}$.  This oscillation correlates with a $\sim$ $10\%$ peak to peak variation of $N_{SrF}^{cell}$ and a $\sim25\%$ peak to peak variation of $N_{SrF}^{beam}$. While temperature-induced variation in the rotational population may explain the variation of $N_{SrF}^{cell}$, it cannot account for the larger variation in $N_{SrF}^{beam}$. We speculate that the background He pressure outside the cell is changing at the 1.4 Hz frequency due to temperature-dependent pumping and/or outgassing rates from the charcoal cryosorb. We have seen that in a similar apparatus cooled with liquid helium rather than a pulse tube, both $N_{SrF}^{cell}$ and $N_{SrF}^{beam}$ vary by 5$\%$ or less shot to shot.

\subsubsection{Source Comparison}

For the production of bright, slow, and cold beams of free radicals and refractory molecular species, this source compares favorably in many respects to competing technologies.  In particular, the brightness $\mathcal{B} = 1.2\times10^{11}$ $N_{rot}=0$ molecules/sr/pulse, is approximately 100 times that produced by a source based on an ablation-seeded room temperature free jet expansion for YbF \cite{Tarbutt2002}.  Furthermore, that free jet expansion beam has a mean forward velocity of 280$\frac{m}{s}$, roughly twice that of the cryogenic buffer gas beam characterized in this paper.  Another group created a beam of SrF by heating SrF$_2\!$ and boron metal to 1550 K \cite{Tu2009}. While the total brightness (over all states) of $2.1 \times 10^{15}$ molecules/sr/s of that source is quite high, the brightness in the rovibrational ground state is $\mathcal{B} = 5 \times 10^{11}$ molecules/sr/s, comparable to the source presented here for $R_{\text{YAG}}= 4$ Hz. However, the high-temperature source can only be operated for a short time before the oven must be refilled. Moreover, the forward (effusive) velocity is $v_{SrF\parallel}^{e} \sim 650 \frac{m}{s}$ at that temperature, undesirable for many experiments.

While the measurements in this work were performed using only SrF, our beam source can be readily adapted by changing the target to create a wide variety of species. Within our group beams of BaF and ThO have been realized using similar techniques, with similar brightness and overall performance.

\section{Conclusion}

We have developed and characterized a robust cryogenic beam source for producing bright, slow beams of translationally and rotationally cold free radicals. We routinely produce a beam with a brightness of $1.2 \times 10^{11}$ $N_{rot}=0$ molecules/sr/pulse in the rovibrational ground state, with forward velocity of $140 \frac{m}{s}$. Under these conditions the source can run for $\gtrsim$ 20 hours before the charcoal cryopumps must be regenerated. For the species SrF we estimate this source can allow $\gtrsim 10^6$ ablation shots, with repetition rates in the range 1-15 Hz, before the source must be opened to change the target. We believe this source may be useful for the wide variety of experiments that require molecular beams of free radical and/or refractory species.

We acknowledge the contributions of N. Hutzler, E. Petrik, D. Patterson, J. Doyle, A. Vutha, P. Orth, M. Steinecker, C. Yale, and C. Bruzewicz. This material is based upon work supported by the ARO, the NSF
and the AFOSR under the MURI award FA9550-09-1-0588.

\bibliography{ColdMoleculesRefs_current}

\end{document}